\documentclass[12pt,number]{elsarticle}
\usepackage{graphicx}
\usepackage{amsmath}
\usepackage{amsfonts}
\usepackage{amssymb}
\usepackage{longtable}

\begin{document}

\journal{Chem. Phys. Lett.}

\begin{frontmatter}
\title{Structure of a low-lying isomer of BOSi$_2$, as a free-space planar
cluster, using the Hartree-Fock method plus second order perturbations}
\author[CTChem]{G. Forte}
\author[CTPhys,SSC,CNISM,INFN]{G. G. N. Angilella\corref{corr}}
\ead{giuseppe.angilella@ct.infn.it}
\author[AntwerpPhys,Oxford]{N. H. March}
\author[CTPhys,CNISM]{R. Pucci}

\address[CTChem]{Dipartimento di Scienze del Farmaco, Universit\`a di Catania,\\
Viale A. Doria, 6, I-95126 Catania, Italy}
\address[CTPhys]{Dipartimento di Fisica e Astronomia, Universit\`a di Catania,\\
Via S. Sofia, 64, I-95123 Catania, Italy}
\address[SSC]{Scuola Superiore di Catania, Universit\`a di Catania,\\ Via
Valdisavoia, 9, I-95123 Catania, Italy}
\address[CNISM]{CNISM, UdR Catania, Via S. Sofia, 64, I-95123 Catania, Italy}
\address[INFN]{INFN, Sez. Catania, Via S. Sofia, 64, I-95123 Catania, Italy}
\address[AntwerpPhys]{Department of Physics, University of Antwerp,\\
Groenenborgerlaan, 171, B-2020 Antwerp, Belgium}
\address[Oxford]{Oxford University, Oxford, UK}
\cortext[corr]{Corresponding author.}

\begin{abstract}

The Hartree-Fock (HF) method, supplemented by low-order M\o{}ller-Plesset (MP2)
perturbation theory, has been utilized to predict the nuclear geometry, assuming
planarity, of a low-lying isomer of the free space cluster BOSi$_2$. The planar
structure found at equilibrium geometry is shown to be stable against small
amplitude molecular vibrations. Finally, some brief comments are made on the
possible relevance of the above free-space cluster geometry to the known B--O
defects which limit the improvement of minority carrier lifetime in a form of
$p$-type silicon.

\medskip
\noindent
PACS: 31.15.Ne, %       Self-consistent-field methods
36.40.Qv%       Stability and fragmentation of clusters
\end{abstract}

\end{frontmatter}

%\section{Introduction}
%\label{sec:introduction}

The main aim of this Letter is to present a theoretical study of the nuclear
geometry of a low-lying isomer of BOSi$_2$ in the form of a free-space planar
cluster. As a neutral cluster, this species has, of course, spin density, and we
have therefore paid attention also to the singly ionized cluster (BOSi$_2$)$^+$.
The effect of such ionization on the equilibrium geometry of the cluster is then
another important focal point. In these calculations, we have utilized the
Hartree-Fock (HF) method, supplemented by second-order M\o{}ller-Plesset (MP2)
perturbation theory \cite{Moeller:34}. Correlation-corrected MP2 calculations
were carried out for the optimizations. All calculations were performed by means
of Gaussian~09 software \cite{Frisch:09a}, where for the augumented
correlation-consistent polarized valence only the quadruple zeta basis set
\cite{Dunning:89} was used (aug-cc-pvqz).

\begin{figure}[t]
\centering
\includegraphics[width=0.8\columnwidth]{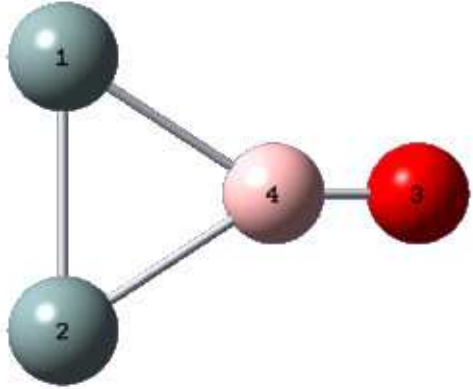}
\caption{(Color online.) Equilibrium structure for the spin compensated, ionized
cluster (BOSi$_2$)$^+$ (1, 2: Si; 3: B; 4: O). We assume planarity and keep the
Si$_2$ separation as in the free-space Si cluster. See Table~\ref{tab:1} for
more structural details.}
\label{fig:1}
\end{figure}

We have taken as starting point a cluster of two free-space Si atoms, held at
first at their equilibrium separation of 2.265~\AA{} \cite{Ragavachari:86}, as
we allow B and O atoms to approach the free-space Si$_2$ complex.
The study of the interaction of the boron oxide subcluster with the silicon dimer
in free space is motivated by the recent interest developed towards the formation
of interstitial (BO)$_n$ impurity complexes in solid-state silicon, which have been considered
in at least two distinct arrangements with respect with the surrounding crystal, and give rise
to different electrical levels and vibrational spectra, depending on their orientation in the crystal
\cite{Carvalho:06}.
Fig.~\ref{fig:1} shows the equilibrium nuclear geometry we predict, when we
restrict ourselves to planarity, for the ionized cluster (BOSi$_2$)$^+$,
assuming spin compensation, and keeping the Si$_2$ separation as in the
free-space Si cluster. The remaining equilibrium bond lengths as calculated by
HF theory are recorded in \AA{} in Table~\ref{tab:1}. The effect of relaxing
then finally the Si$_2$ separation to achieve minimum energy is shown in
Table~\ref{tab:1}, the difference in energy due to such final relaxation being
extremely small, as shown in Table~\ref{tab:1}.

Table~\ref{tab:2} then records the final relaxed cluster geometry for the
neutral case of BOSi$_2$, again, however, with planarity assumed. The relatively
small changes in the equilibrium bond lengths due to `charging' is clear from
comparison between Tables~\ref{tab:1} and \ref{tab:2}.

\begin{figure}[t]
\centering
\includegraphics[width=0.4\columnwidth]{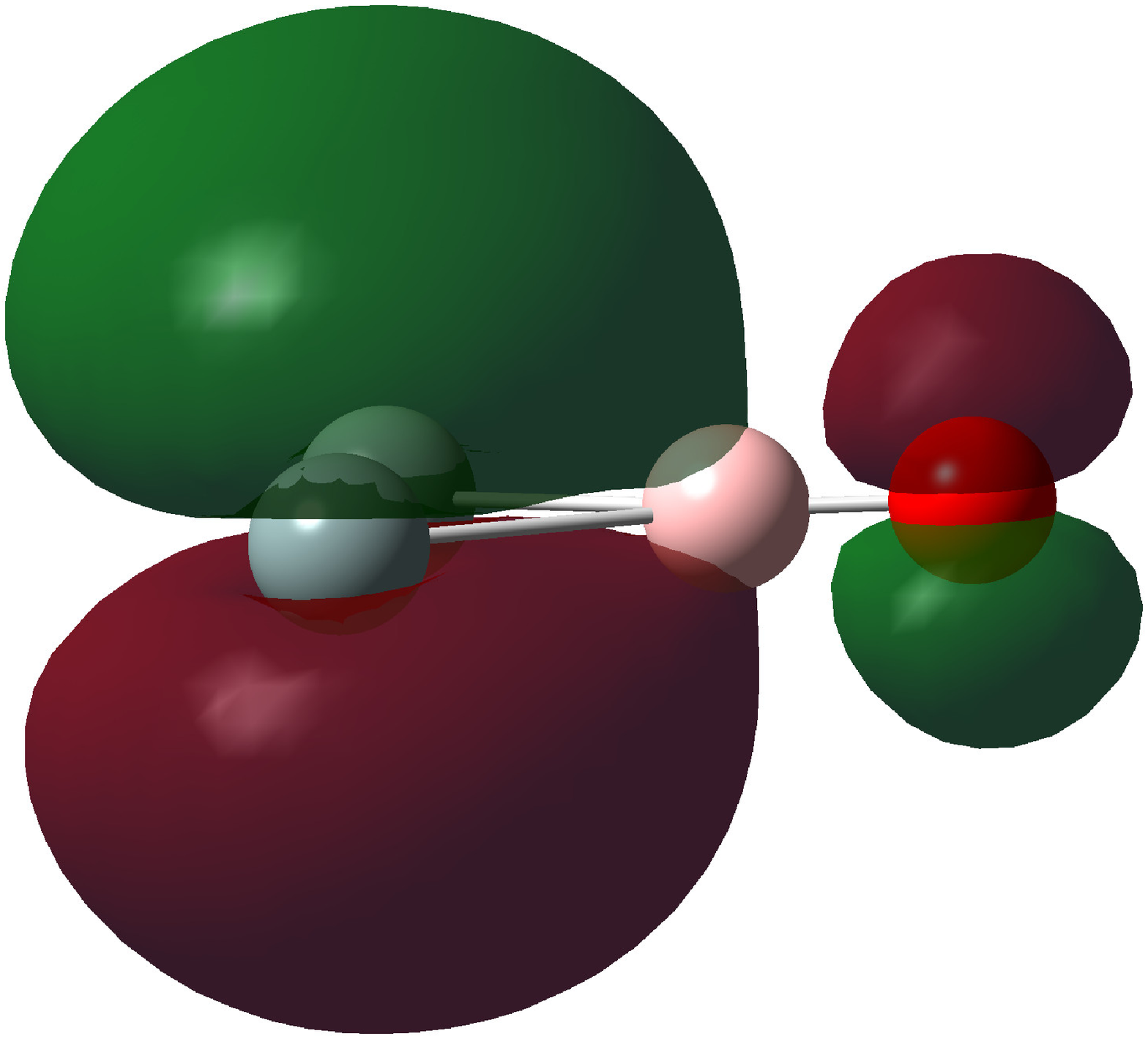}
\includegraphics[width=0.4\columnwidth]{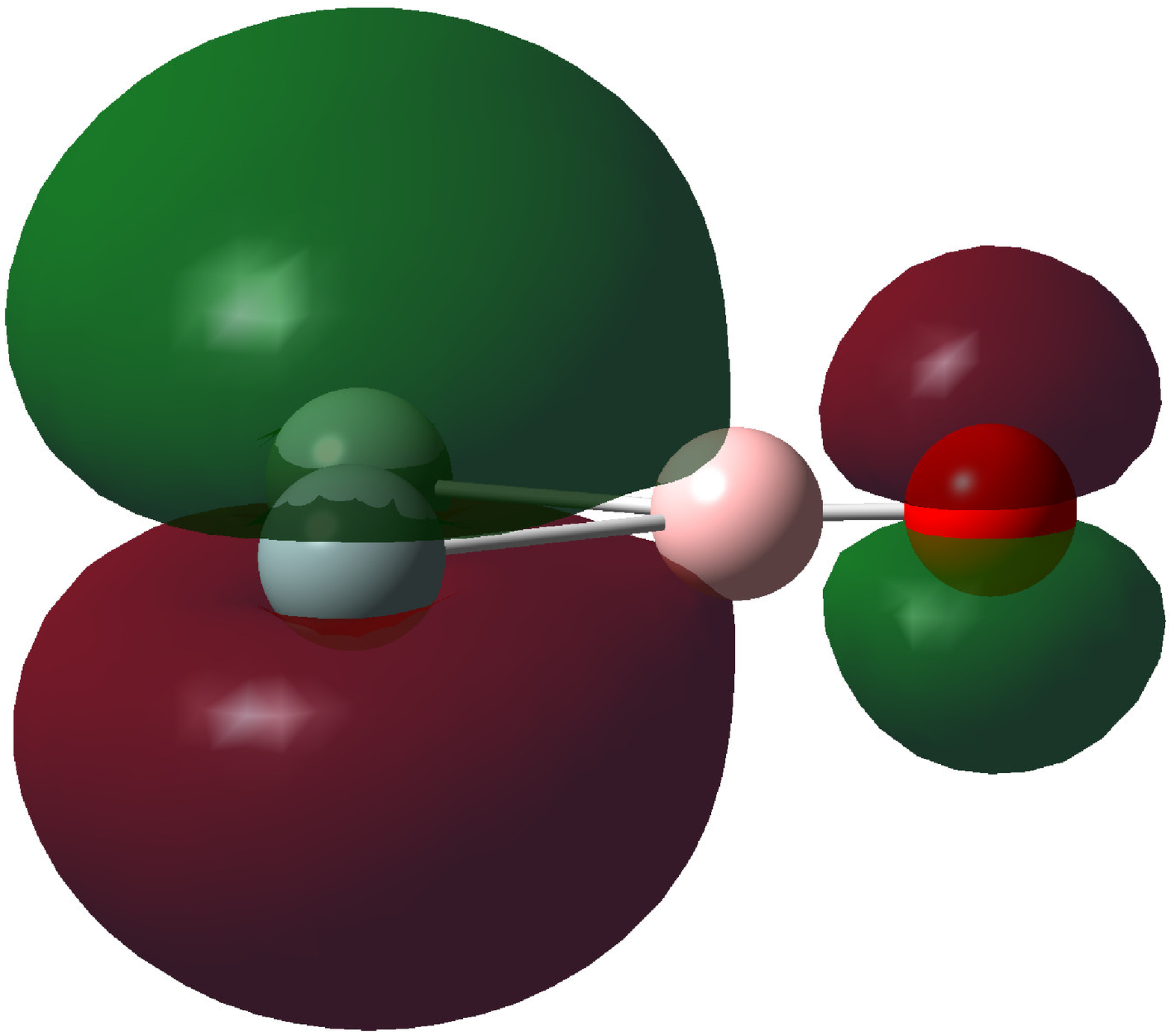}
\caption{(Color online.) HOMO orbitals of the neutral (left panel) and charged cluster (right panel).}
\label{fig:2}
\end{figure}

Tables~\ref{tab:1} and \ref{tab:2} show, together with Fig.~\ref{fig:1}, the
rather remarkable property that the atoms 1, 2, and 4 form an almost equilateral
triangle, with the O atom bonded directly to B, with bond length 1.22~\AA. In
this context, we also record in the two Tables the dipole moments for both
neutral and ionized clusers. Fig.~\ref{fig:2} shows the HOMO orbitals of the two clusters, while Table~\ref{tab:NBO} reports the NBO charges of the two clusters.

\begin{figure}[t]
\centering
\includegraphics[bb=530 151 118 695,clip,height=0.45\columnwidth,angle=-90]{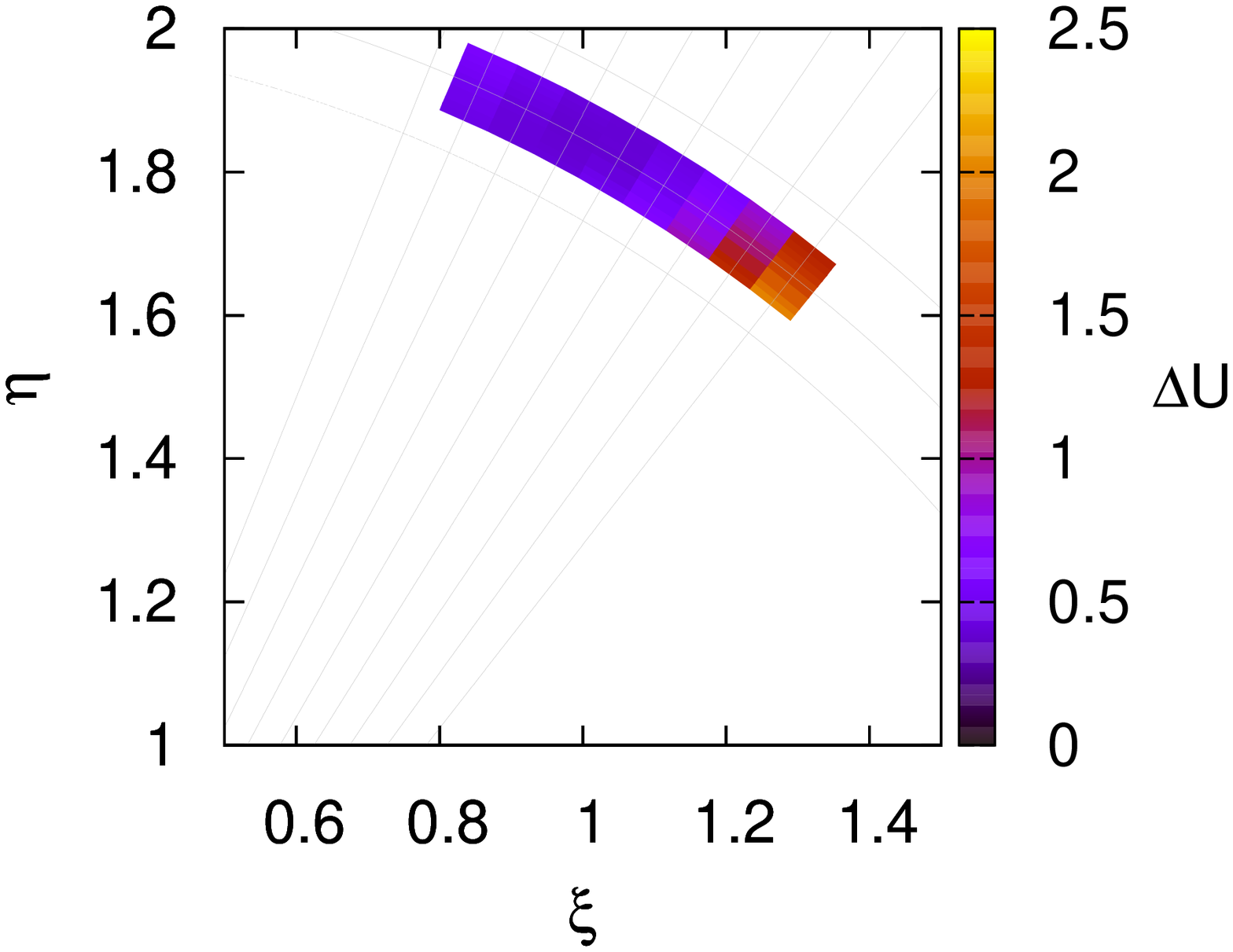}
\includegraphics[bb=530 151 118 695,clip,height=0.45\columnwidth,angle=-90]{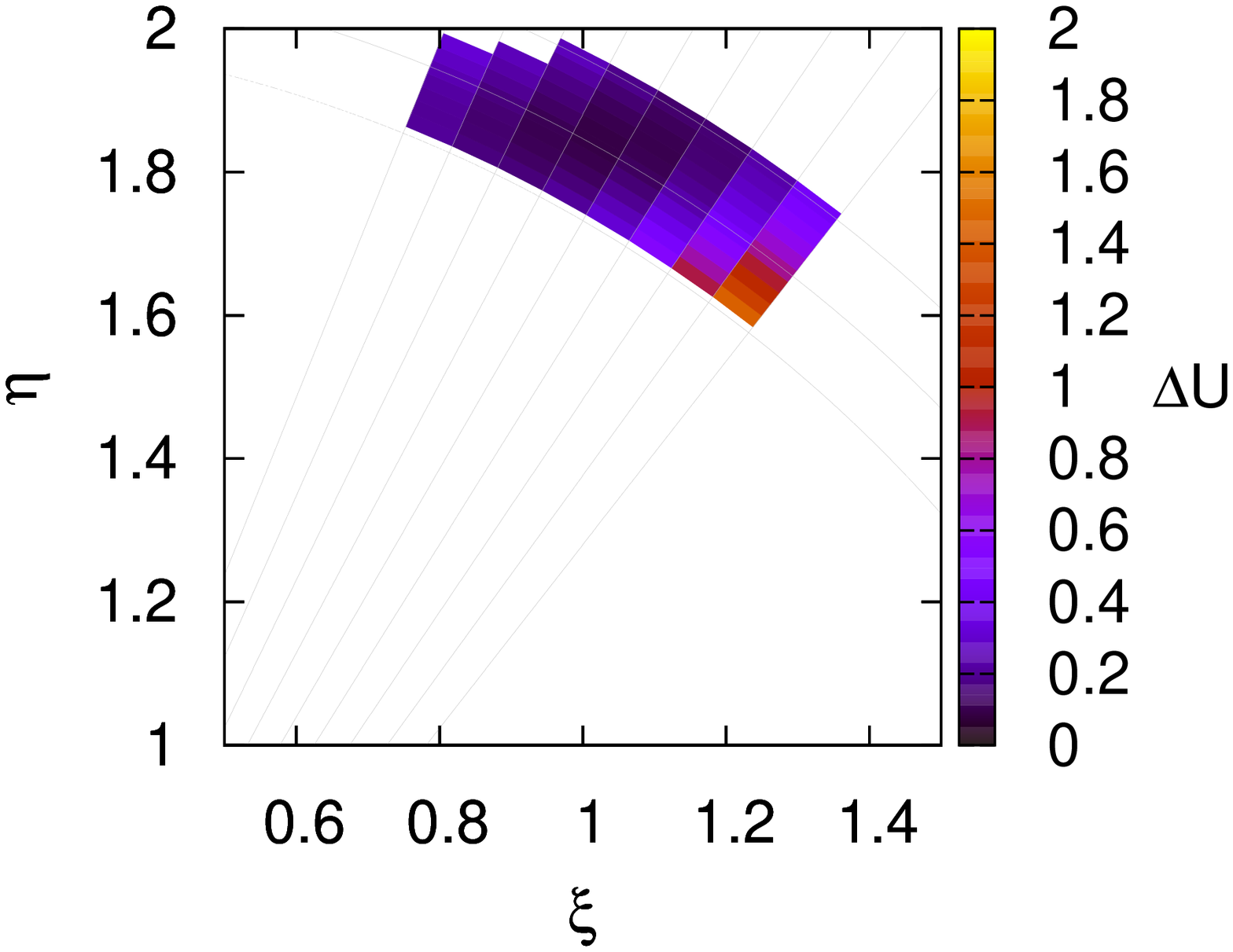}
\caption{(Color online.) Contour plots of the potential energy surfaces
for neutral BOSi$_2$ (left panel) and singly ionized (BOSi$_2$)$^+$ (right panel),
as a function of $\xi = R\cos\theta$ and $\eta=R\sin\theta$, where
$R=2.0-2.2$~\AA{} is the Si--B distance, and $\theta=52^\circ -68^\circ$ is the $\angle$Si,B,Si angle.
Actually plotted is the difference $\Delta U$ of the potential energy with
respect to its minimum value, in each case.
Distances are in \AA, while energies are here in eV.}
\label{fig:3}
\end{figure}

Potential energy surfaces (PES) were evaluated for both 
neutral BOSi$_2$ and singly ionized (BOSi$_2$)$^+$ (Fig.~\ref{fig:3}).
To this aim,
we have maintanied the same basis set as in the structural calculations,
but applying the PBE functional for the neutral cluster,
while PBE1 was adopted for the charged system,
since convergence criterions are not met with PBE.
We have explored the PES in a range of the Si--B distance $R=2.06-2.18$~\AA,
with the angle $\theta=\angle\mathrm{Si,B,Si}=52-66^\circ$.
Fig.~\ref{fig:3} shows that, in strong
agreement with MP2 calculations, absolute minima are found for $R\simeq 2.10$~\AA,
$\theta\simeq 60^\circ$ for the neutral
cluster, and $R\simeq 2.12$~\AA{} and $\theta\simeq 60^\circ$ for (BOSi$_2$)$^+$.

We have also studied the frequencies of the normal modes of vibration for both
the charged and neutral clusters (Table~\ref{tab:freq}). All the frequencies are
real, which implies the stability of both clusters against small vibrations.

Finally, for the neutral planar cluster BOSi$_2$, with its structure as recorded
in Table~\ref{tab:2}, the cluster energy for this low-lying isomer is recorded
in Table~\ref{tab:freq}. The energy required to separate the above planar
clusters into its free-space neutral atoms is of 0.5981~Hartree for the neutral
cluster, and of 0.5663~Hartree for the singly ionized cluster.

In relation to the above predictions, it would, of course, be interesting if
experiments could be carried out on such clusters of BOSi$_2$ as are studied
here, to ascertain whether our predicted planar nuclear geometries are among the
low-lying isomers of such species.

But, in the absence, to our knowledge, of such experiments, we shall have
recourse below to some recent theoretical work by Chen \emph{et al.}
\cite{Chen:13}, which, however, deals with the structure of boron-oxygen defects
in a form of $p$-type silicon crystal (see also \cite{Carvalho:06}). 

To summarize briefly, the main achievements of the present theoretical study are
the HF-MP2 bond lengths shown for the planar clusters of BOSi$_2$ in
Tables~\ref{tab:1} and \ref{tab:2}, corresponding to the singly ionized and
neutral free-space clusters, respectively. While no experimental results on such
free-space clusters seem to be available presently, some contact can be made
with available theoretical work on BO defects in a particular $p$-type silicon
crystal. In the latter solid-state assembly, the Si--B distance of 1.90~\AA{} is
shown in Fig.~1(a) of Ref.~\cite{Carvalho:06}, while 2.07~\AA{} is recorded in
their Fig.~1(b). Our Table~\ref{tab:2} predicts the cluster Si--B length as
2.1~\AA, which is already in excellent accord with the above values.

As mentioned earlier, we believe, in view of our present predictions, that
experiments on free-space clusters BOSi$_2$ would now be of interest.

\section*{Acknowledgements}

NHM wishes to acknowledge that his contribution to the present article was made
during a visit to the University of Catania. NHM thanks Professors R. Pucci and
G. G. N. Angilella for their kind hospitality.

\bibliographystyle{rsc}
\bibliography{a,b,c,d,e,f,g,h,i,j,k,l,m,n,o,p,q,r,s,t,u,v,w,x,y,z,zzproceedings,Angilella}

\providecommand*{\mcitethebibliography}{\thebibliography}
\csname @ifundefined\endcsname{endmcitethebibliography}
{\let\endmcitethebibliography\endthebibliography}{}
\begin{mcitethebibliography}{6}
\providecommand*{\natexlab}[1]{#1}
\providecommand*{\mciteSetBstSublistMode}[1]{}
\providecommand*{\mciteSetBstMaxWidthForm}[2]{}
\providecommand*{\mciteBstWouldAddEndPuncttrue}
  {\def\EndOfBibitem{\unskip.}}
\providecommand*{\mciteBstWouldAddEndPunctfalse}
  {\let\EndOfBibitem\relax}
\providecommand*{\mciteSetBstMidEndSepPunct}[3]{}
\providecommand*{\mciteSetBstSublistLabelBeginEnd}[3]{}
\providecommand*{\EndOfBibitem}{}
\mciteSetBstSublistMode{f}
\mciteSetBstMaxWidthForm{subitem}
{(\emph{\alph{mcitesubitemcount}})}
\mciteSetBstSublistLabelBeginEnd{\mcitemaxwidthsubitemform\space}
{\relax}{\relax}

\bibitem[M\o{}ller and Plesset(1934)]{Moeller:34}
C.~M\o{}ller and M.~S. Plesset, \emph{Phys. Rev.}, 1934, \textbf{46}, 618\relax
\mciteBstWouldAddEndPuncttrue
\mciteSetBstMidEndSepPunct{\mcitedefaultmidpunct}
{\mcitedefaultendpunct}{\mcitedefaultseppunct}\relax
\EndOfBibitem
\bibitem[Frisch \emph{et~al.}(2009)Frisch, Trucks, Schlegel, Scuseria, Robb,
  Cheeseman, Scalmani, Barone, Mennucci, Petersson, Nakatsuji, Caricato, Li,
  Hratchian, Izmaylov, Bloino, Zheng, Sonnenberg, Hada, Ehara, Toyota, Fukuda,
  Hasegawa, Ishida, Nakajima, Honda, Kitao, Nakai, Vreven, Montgomery, Jr.,
  Peralta, Ogliaro, Bearpark, Heyd, Brothers, Kudin, Staroverov, Kobayashi,
  Normand, Raghavachari, Rendell, Burant, Iyengar, Tomasi, Cossi, Rega, Millam,
  Klene, Knox, Cross, Bakken, Adamo, Jaramillo, Gomperts, Stratmann, Yazyev,
  Austin, Cammi, Pomelli, Ochterski, Martin, Morokuma, Zakrzewski, Voth,
  Salvador, Dannenberg, Dapprich, Daniels, Farkas, Foresman, Ortiz, Cioslowski,
  and Fox]{Frisch:09a}
M.~J. Frisch, G.~W. Trucks, H.~B. Schlegel, G.~E. Scuseria, M.~A. Robb, J.~R.
  Cheeseman, G.~Scalmani, V.~Barone, B.~Mennucci, G.~A. Petersson,
  H.~Nakatsuji, M.~Caricato, X.~Li, H.~P. Hratchian, A.~F. Izmaylov, J.~Bloino,
  G.~Zheng, J.~L. Sonnenberg, M.~Hada, M.~Ehara, K.~Toyota, R.~Fukuda,
  J.~Hasegawa, M.~Ishida, T.~Nakajima, Y.~Honda, O.~Kitao, H.~Nakai, T.~Vreven,
  J.~A. Montgomery, Jr., J.~E. Peralta, F.~Ogliaro, M.~Bearpark, J.~J. Heyd,
  E.~Brothers, K.~N. Kudin, V.~N. Staroverov, R.~Kobayashi, J.~Normand,
  K.~Raghavachari, A.~Rendell, J.~C. Burant, S.~S. Iyengar, J.~Tomasi,
  M.~Cossi, N.~Rega, J.~M. Millam, M.~Klene, J.~E. Knox, J.~B. Cross,
  V.~Bakken, C.~Adamo, J.~Jaramillo, R.~Gomperts, R.~E. Stratmann, O.~Yazyev,
  A.~J. Austin, R.~Cammi, C.~Pomelli, J.~W. Ochterski, R.~L. Martin,
  K.~Morokuma, V.~G. Zakrzewski, G.~A. Voth, P.~Salvador, J.~J. Dannenberg,
  S.~Dapprich, A.~D. Daniels, O.~Farkas, J.~B. Foresman, J.~V. Ortiz,
  J.~Cioslowski and D.~J. Fox, \emph{Gaussian 09, \uppercase{R}evision
  \uppercase{A}.02}, Gaussian, Inc., Wallingford, CT, 2009\relax
\mciteBstWouldAddEndPuncttrue
\mciteSetBstMidEndSepPunct{\mcitedefaultmidpunct}
{\mcitedefaultendpunct}{\mcitedefaultseppunct}\relax
\EndOfBibitem
\bibitem[{Dunning, Jr.}(1989)]{Dunning:89}
T.~H. {Dunning, Jr.}, \emph{J. Chem. Phys.}, 1989, \textbf{90}, 1007\relax
\mciteBstWouldAddEndPuncttrue
\mciteSetBstMidEndSepPunct{\mcitedefaultmidpunct}
{\mcitedefaultendpunct}{\mcitedefaultseppunct}\relax
\EndOfBibitem
\bibitem[Raghavachari(1986)]{Ragavachari:86}
K.~Raghavachari, \emph{J. Chem. Phys.}, 1986, \textbf{84}, 5672\relax
\mciteBstWouldAddEndPuncttrue
\mciteSetBstMidEndSepPunct{\mcitedefaultmidpunct}
{\mcitedefaultendpunct}{\mcitedefaultseppunct}\relax
\EndOfBibitem
\bibitem[Carvalho \emph{et~al.}(2006)Carvalho, Jones, Sanati, Estreicher,
  Coutinho, and Briddon]{Carvalho:06}
A.~Carvalho, R.~Jones, M.~Sanati, S.~K. Estreicher, J.~Coutinho and P.~R.
  Briddon, \emph{Phys. Rev. B}, 2006, \textbf{73}, 245210\relax
\mciteBstWouldAddEndPuncttrue
\mciteSetBstMidEndSepPunct{\mcitedefaultmidpunct}
{\mcitedefaultendpunct}{\mcitedefaultseppunct}\relax
\EndOfBibitem
\bibitem[Chen \emph{et~al.}(2013)Chen, Yu, Zhu, Chen, and Yang]{Chen:13}
X.~Chen, X.~Yu, X.~Zhu, P.~Chen and D.~Yang, \emph{Applied Physics Express},
  2013, \textbf{6}, 041301\relax
\mciteBstWouldAddEndPuncttrue
\mciteSetBstMidEndSepPunct{\mcitedefaultmidpunct}
{\mcitedefaultendpunct}{\mcitedefaultseppunct}\relax
\EndOfBibitem
\end{mcitethebibliography}

\clearpage

\begin{table}
\centering
\begin{tabular}{lcc}
\hline\hline
 & Si--Si free & Si--Si fixed \\
\hline
$E$ (a.u.)             & $-677.7901$  & $-677.7847$ \\
dipole (Debye)		   & 6.38	  & 6.66 \\ 
symmetry			   & $C_{2v}$ &       $C_{2v}$ \\
1--2 (\AA)			   & 2.11	  & 2.26 \\
1--4 (\AA)			   & 2.11	  & 2.13 \\
2--4 (\AA)			   & 2.11	  & 2.13 \\
3--4 (\AA)			   & 1.22	  & 1.22 \\
$\angle$1,4,2 ($^\circ$) & 59.83	  & 64.08 \\
$\angle$1,2,4 ($^\circ$) & 60.09	  & 57.96 \\
\hline\hline
\end{tabular}
\caption{Predicted structural details for (BOSi$_2$)$^+$ in the equilibrium
geometry, as shown in Fig.~\ref{fig:1}. Both columns refer to a planar geometry,
but with free Si--Si distance (first column), and Si--Si fixed as in the
free-space Si$_2$ cluster (second column), \emph{i.e.} 2.2648~\AA{}
\cite{Ragavachari:86}.}
\label{tab:1}
\end{table}

\begin{table}
\centering
\begin{tabular}{lcc}
\hline\hline
 & Si--Si free & Si--Si fixed \\
\hline
$E$ (a.u.)                  &  $-678.1352$ &   $-678.1273$ \\
dipole (Debye)		    &  4.41 	 & 4.3 \\
symmetry			    &  $C_{2v}$	 & 	   $C_{2v}$ \\
1--2 (\AA)			    &  2.12 	 & 2.26 \\
1--4 (\AA)			    &  2.1  	 & 2.11 \\
2--4 (\AA)			    &  2.1  	 & 2.11 \\
3--4 (\AA)			    &  1.22 	 & 1.23 \\
$\angle$1,4,2 ($^\circ$)  &  60.71	 & 64.73 \\
$\angle$1,2,4 ($^\circ$)  &  59.64	 & 57.64 \\
\hline\hline
\end{tabular}
\caption{Predicted structural details for neutral cluster BOSi$_2$ in the
equilibrium planar geometry.
The energies of the isolated atoms are 
$-288.961$~a.u. (Si), $-75.0056$~a.u. (O), $-24.6016$~a.u. (B), yielding
a total energy difference of $0.5981$~a.u. of the isolated atoms, with respect to
the ground-state energy of the neutral cluster BOSi$_2$ (estimated with fixed Si--Si
distance).}
\label{tab:2}
\end{table}

\begin{table}
\centering
\begin{tabular}{cr@{.}lr@{.}l}
\hline\hline
$i$ & \multicolumn{2}{c}{BOSi$_2$} & \multicolumn{2}{c}{(BOSi$_2$)$^+$} \\
\hline
1 &	0 & 238 & 0 & 755 \\
2 &	0 & 238 & 0 & 755 \\
4 &	0 & 412 & 0 & 234 \\
3 &	$-0$ & 888 & $-0$ & 743 \\
\hline\hline
\end{tabular}
\caption{NBO charges on the $i$th atom of neutral and
singly ionized BOSi$_2$ cluster, with free Si--Si distance.}
\label{tab:NBO}
\end{table}

\begin{table}
\centering
\begin{tabular}{r@{.}lr@{.}l}
\hline\hline
\multicolumn{2}{c}{BOSi$_2$} & \multicolumn{2}{c}{(BOSi$_2$)$^+$} \\
\hline
180 & 89 & 140 & 75 \\
338 & 13 & 385 & 18 \\
394 & 85 & 387 & 96$^\ast$ \\
584 & 95 & 391 & 41 \\
628 & 56$^\ast$ & 559 & 93 \\
1793 & 73 & 1819 & 81 \\
\hline\hline
\end{tabular}
\caption{Vibrational frequencies (in cm$^{-1}$) of normal modes of neutral and
singly ionized BOSi$_2$ cluster, with free Si--Si distance. In particular, frequencies labelled with $^\ast$ refer to out-of-plane normal modes.}
\label{tab:freq}
\end{table}

\end{document}